# Heat Conduction and Charge Ordering in Perovskite Manganites, Nickelates and Cuprates

**ABSTRACT**

We discuss heat transport in hole-doped manganites, nickelates and cuprates for which real-space charge ordering has been observed. We survey the thermal transport response to charge order in the various materials and the associated structural modifications, particularly distortions of the metal-oxygen polyhedra associated with localized charge, that may be a principal source of phonon damping.

**INTRODUCTION**

The real-space ordering of doped holes (and spins) in perovskite manganites, nickelates and cuprates has attracted considerable attention recently. These phenomena reflect a novel interplay between charge, spin and lattice degrees of freedom. In the cuprates the charge order, when static and long-ranged, suppresses superconductivity [1]; there is evidence to suggest that fluctuating charge and spin order in the cuprates is necessary for high-temperature superconductivity [2]. In all of the title compounds the lattice heat conduction is predominant and thus the study of thermal conductivity affords an opportunity to investigate the effects of the ordering phenomena on the damping of long-wavelength phonons.

This paper reviews the relevant heat transport data in these materials, emphasizing the empirical connection between phonon damping and structural distortions related to localized charge. The charge ordering in these compounds is closely associated with antiferromagnetic ordering of the metal-ion spins. However, the heat transport is rather insensitive to the spin order, and we omit the latter from our discussion.

---
Joshua L. Cohn, Department of Physics, University of Miami, Coral Gables, FL 33124-0530

# MANGANITES

Doped perovskite manganites [3] $R_{1-x}A_xMnO_3$ (R=La, Pr, Nd, Sm; A=Ca, Sr, Ba, Pb), exhibit a complex phase behavior [Fig. 1 (a)] and rich physics. Though our main focus in the present work is on compositions for which charge and orbital ordering occur, we briefly discuss our studies of heat transport over a broader range of doping [4] since they serve as the motivation for our hypothesis that thermal transport is a sensitive probe of the local structure in the metal-oxide perovskites generally. The parent compound, $RMnO_3$ ($Mn^{3+}$; $t_{2g}^3 e_g^1$) is an antiferromagnetic insulator. The $Mn^{3+}O_6$ octahedra undergo a Jahn-Teller deformation that splits the degeneracy of the two $e_g$ states and yields two elongated bonds corresponding to the $d_{z^2}$ orbitals. Divalent substitution for $R^{3+}$, in the simplest picture, converts x $Mn^{3+}$ ions to $Mn^{4+}$. The latter is not Jahn-Teller active, thus the average $MnO_6$ distortion

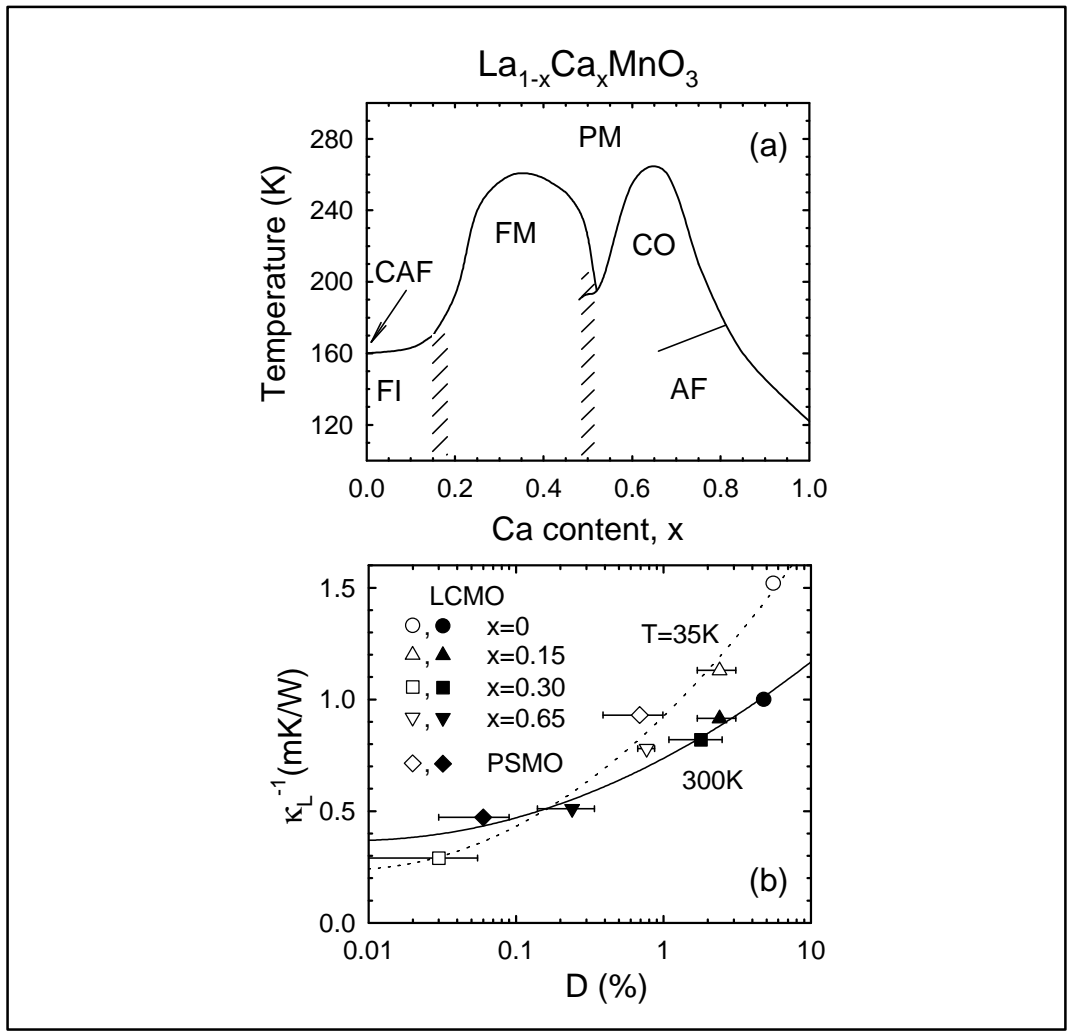

FIGURE 1 (a) Phase diagram for LCMO. Labels indicate the paramagnetic insulating (PM), canted antiferromagnetic insulating (CAF), ferromagnetic insulating (FI), ferromagnetic metallic (FM), charge-ordered insulating (CO) and antiferromagnetic insulating (AF) states. (b) Thermal resistivity *vs* $MnO_6$ distortion (see text) for LCMO and PSMO polycrystals [4].

decreases with increasing x. Charge transport may be viewed as occurring by transfer of $e_g$ holes on $Mn^{4+}$ to $e_g$ states on neighboring $Mn^{3+}$, and occurs via thermally activated hopping of small polarons in the high-temperature paramagnetic (PM) phase.

The undoped compound has a thermal conductivity $\kappa \sim 1$ W/mK at 300 K, comparable to the theoretical minimum value [5], and consistent with a high degree of disorder attributable to the Jahn-Teller distortions. Figure 1 (b) shows that the lattice thermal resistivity, $W_L=1/\kappa_L$, for all compositions scales with the bond disorder, defined from neutron diffraction measurements as, $D \equiv (1/3)\Sigma|u_i-\underline{u}|/\underline{u} \times 100$, with $u_i$ the Mn-O bond lengths, and $\underline{u}=(u_1u_2u_3)^{1/3}$. This correlation is especially compelling because it holds at both high-T and low-T while D is dramatically altered by the ferromagnetic and charge-ordering transitions upon cooling from the PM state. For example, $Pr_{0.5}Sr_{0.5}MnO_3$ (PSMO) has the smallest D at 300 K, but one of the largest at 35K; the reverse is true for $La_{0.7}Ca_{0.3}MnO_3$. The various phase transitions all involve modifications of the local structure [6] which correlate with hole itinerancy and the magnetism, consistent with a polaronic origin for the lattice distortions.

Consider the data for $Pr_{0.5}Sr_{0.5}MnO_3$ and $La_{0.35}Ca_{0.65}MnO_3$ in more detail near their FM-CO and PM-CO transitions, respectively (Fig. 2). First note that the additional thermal resistivity that develops in the CO phases (relative to that at $1.2T_{CO}$), computed using the solid curves in Fig. 2 (a) and (b), follows a mean-field

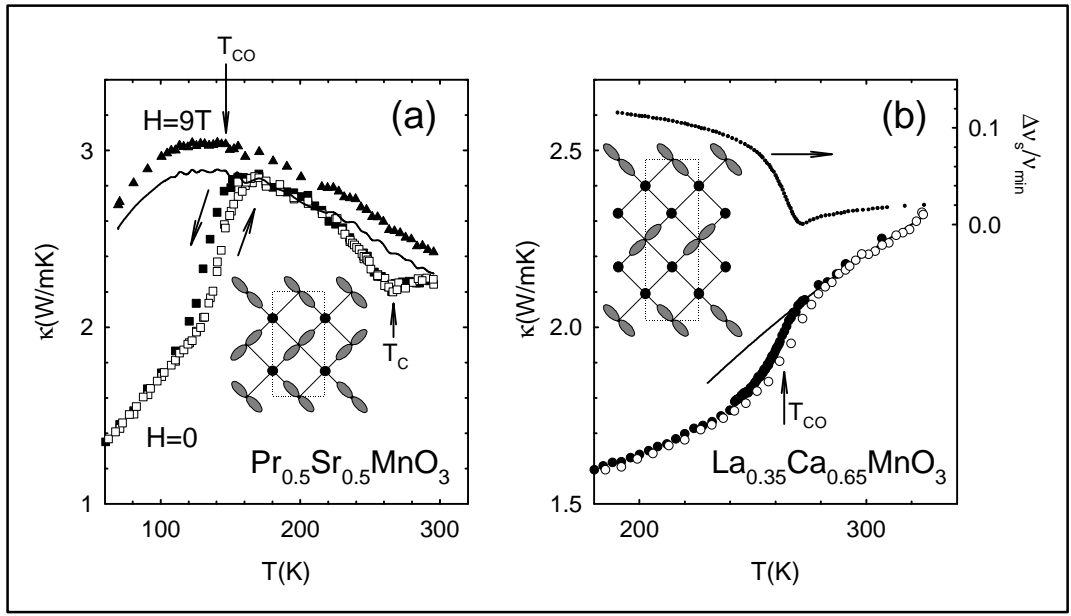

FIGURE 2 (a) Thermal conductivity *vs* temperature for (a) $Pr_{0.5}Sr_{0.5}MnO_3$ and (b) $La_{0.35}Ca_{0.65}MnO_3$ [4] along with the relative change in sound velocity [9] for the latter. The solid lines represent in (a) the data at H=9T, scaled to match the data above $T_{CO}=135K$, and in (b) an extrapolated polynomial fit to the data at $T>T_{CO}=265$ K. The insets in (a) and (b) represent the charge and orbital ordering pattern in the ac-planes below $T_{CO}$ [7,8], with doubling and tripling of the unit cells along the a-axis, respectively (dotted rectangles): $Mn^{4+}$ (solid circles), $Mn^{3+}$ (hatched lobes, representing $d_{z^2}$-orbitals).

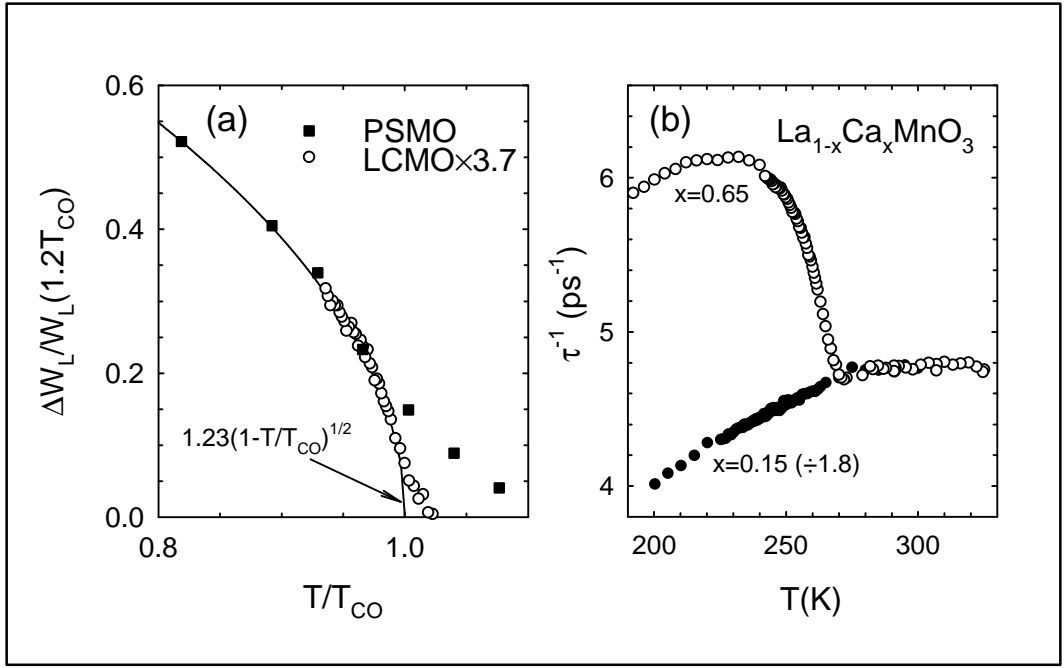

FIGURE 3 (a) Normalized change in thermal resistivity from that extrapolated from $T>T_{CO}$ for data from Fig. 2 plotted *vs* reduced temperature. (b) phonon scattering rate *vs* temperature for $La_{1-x}Ca_xMnO_3$ x=0.15, 0.65, computed as described in the text.

behavior [Fig. 3 (a)], $\Delta W_L/W_L \propto (1-T/T_{CO})^{1/2}$, i.e. scales with the structural order parameter for the CO phase, e.g. the integrated intensity of the charge-ordered superlattice reflections from x-ray diffraction studies of a specimen of similar composition [7]. The ordering pattern of the holes and $Mn^{3+}$ $d_{z^2}$ orbitals is depicted for both compounds in the insets of Fig. 2 [7,8]. A fundamental question for our studies of heat transport in all of the perovskite materials is how charge ordering influences the damping of heat-carrying phonons. To address this issue it is useful to consider the kinetic theory expression for the lattice thermal resistivity, $W_L=(3/C_L v^2)\tau^{-1}$, where $C_L$ is the lattice specific heat, $v$ is the sound velocity, and $\tau^{-1}$ is the phonon relaxation rate. Fig. 2 (b) shows that in the case of the manganites, the charge ordered phase is characterized by a substantial hardening of the lattice [9], with $v$ increasing by 10% just below $T_{CO}$. The background lattice specific heat is continuous well above and below $T_{CO}$ [7], and thus the $\kappa$ data imply a substantial increase in phonon damping ($\tau^{-1}$) below $T_{CO}$ (the volume change is negligible [7]). Figure 3 (b) shows $\tau^{-1}(T)$ computed for $La_{0.35}Ca_{0.65}MnO_3$ from the $\kappa$ data using $C_L(T)$ and $v(T)$ from Ref. 9 and taking $v(T_{CO})=3,000$ m/s. Also shown is $\tau^{-1}(T)$ for $La_{0.85}Ca_{0.15}MnO_3$, computed assuming $v=3,000$ m/s independent of T, and scaled to match the data for x=0.65 at $T>T_{CO}$. The x=0.15 compound remains in the PM phase for T>140 K, and thus serves as a reference for the phonon scattering rate in the absence of charge ordering. $\tau^{-1}$ for x=0.65 increases abruptly by nearly 40% just below the transition, and then decreases at T<230 K, paralleling the scaled $\tau^{-1}(T)$ for the x=0.15 compound. This constant offset of the $\tau^{-1}$ curves indicates that the enhanced scattering in the CO phase is associated with *static bond disorder*.

It is significant that $\tau^{-1}$ for $La_{0.35}Ca_{0.65}MnO_3$ shows no anomalous behavior at $T>T_{CO}$ whereas $v(T)$ indicates a softening of the lattice, presumably associated with charge-order fluctuations as $T_{CO}$ is approached from above. Evidently such *dynamical bond disorder* (presumably with a frequency related to that of an optical phonon involved in polaron hopping) does not necessarily yield damping of heat-carrying phonons.

**NICKELATES**

The doped holes in $La_{2-x}Sr_xNiO_{4+\delta}$ [10] order two-dimensionally into periodic, quasi-one-dimensional stripes within the $NiO_2$ planes. This effect is most pronounced for the hole concentration x=1/3 where the charge-stripe period is commensurate with the lattice [Fig. 4 (a)].

Recently, Hess *et al*. [11] reported thermal conductivity measurements for these compounds [Fig. 4 (b)]. In contrast to $La_{0.35}Ca_{0.65}MnO_3$, $\kappa(T)$ for $La_{5/3}Sr_{1/3}NiO_4$ *increases* below $T_{CO}$=240 K. The sound velocity [12] is enhanced in the charge-ordered phase, but the effect is two orders of magnitude smaller than that in the manganite. Therefore $\tau^{-1}$ *decreases* substantially for the nickelate within the CO phase. Phononic Raman scattering [13] indicates broad phonon modes at $T>T_{CO}$, indicative of polaronic effects and an inhomogeneous charge distribution.

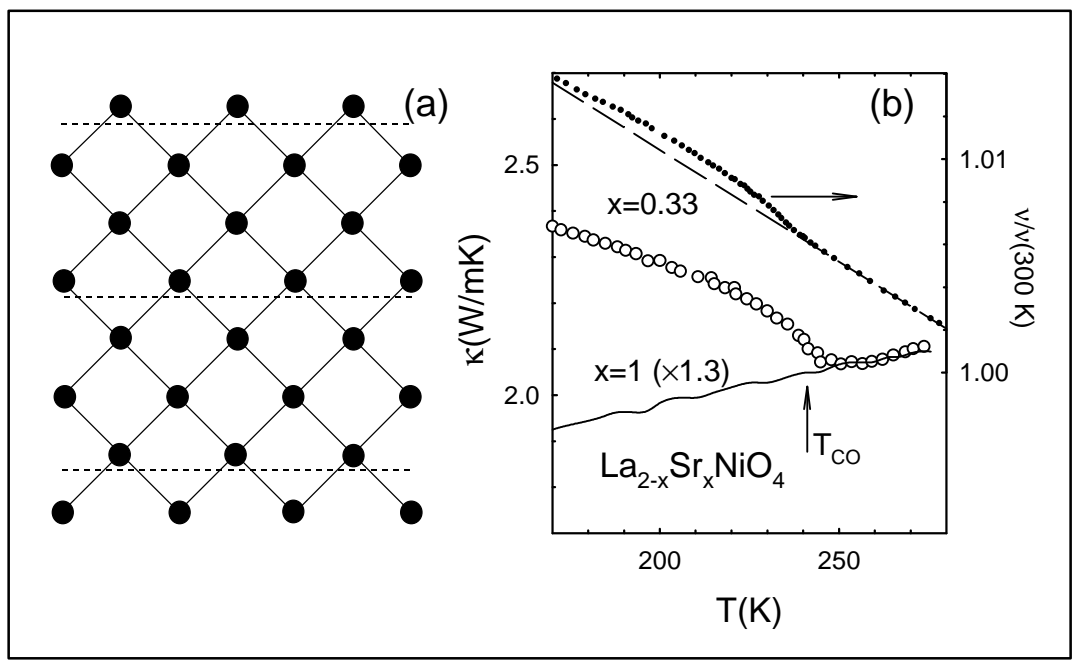

FIGURE 4 (a) Charge ordering scheme within a $NiO_2$ plane for $La_{5/3}Sr_{1/3}NiO_4$: solid circles and lines represent the Ni lattice, dashed lines the charge stripes (centered between Ni and O atoms). (b) Thermal conductivity [11] and sound velocity [12] *vs* temperature for $La_{2-x}Sr_xNiO_4$ polycrystals. The data for x=1 has been scaled to match the data of x=0.33 at $T>T_{CO}\approx 240$ K.

Based on our observations for the manganites it is plausible that the suppressed phonon scattering in the nickelate is associated with a decrease in the average octahedral distortion in the stripe-ordered phase.

## CUPRATES

### (La,Nd)$_{2-x}$Sr$_x$CuO$_4$

Rare-earth substitution for La ions in La$_{2-x}$Sr$_x$CuO$_4$ (LSCO) induces a structural transition from the low-temperature orthorhombic (LTO) to low-temperature tetragonal (LTT) phase at low temperatures. The transition involves a rotation of the tilt axis of the CuO$_6$ octahedra from the [100] direction (at 45 to the Cu-O bonds) to [110] (along the Cu-O bonds), with no change in the magnitude of the tilts (see Fig. 5). The [110] tilts are effective in pinning charge stripes oriented along the Cu-O bonds and stabilize static, long-ranged charge stripe order which suppresses superconductivity [14]. For sufficiently small tilt angles, the LTT phase is superconducting, evidently because their pinning potential is lower [15].

Baberski *et al.* [16] studied thermal conductivity in a series of Nd- and Eu-doped LSCO polycrystals. They found an abrupt jump in $\kappa$ at the LTO-LTT transition [Fig. 6 (a)] *but only when the LTT phase was nonsuperconducting*. The authors concluded that the anomaly in the heat transport was caused by the scattering of phonons by fluctuating charge stripes in the LTO phase, and suppression of this scattering for static stripes. For fixed RE content, the size of the anomaly, $\Delta\kappa/\kappa$, was observed to decrease gradually with increasing Sr composition,

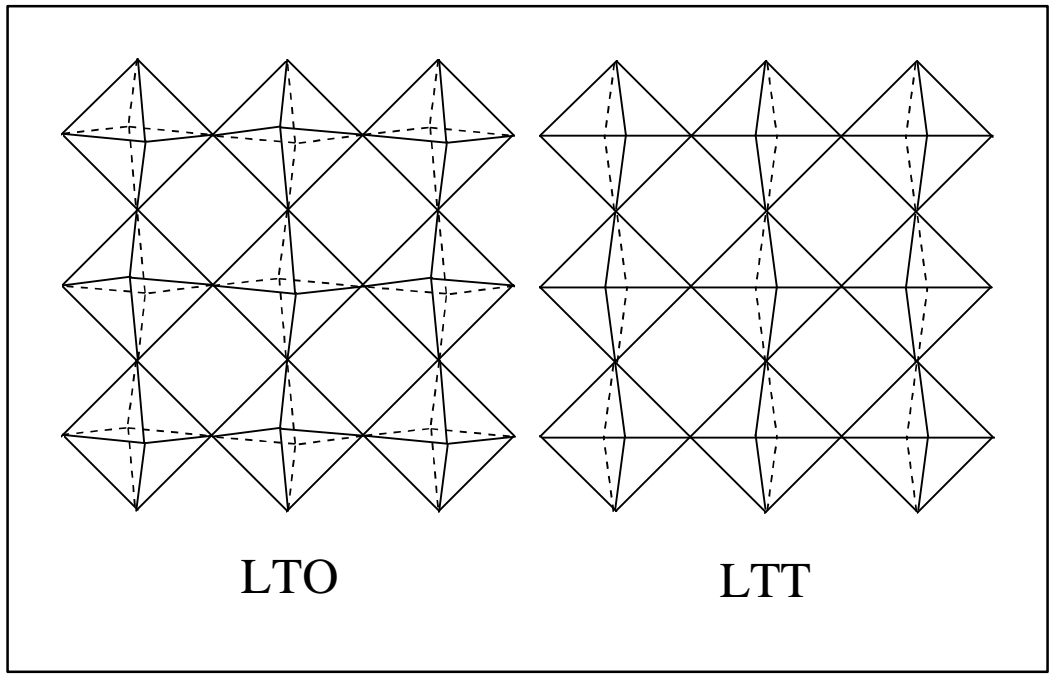

FIGURE 5 Tilt pattern of CuO$_6$ octahedra in the LTO and LTT phases.

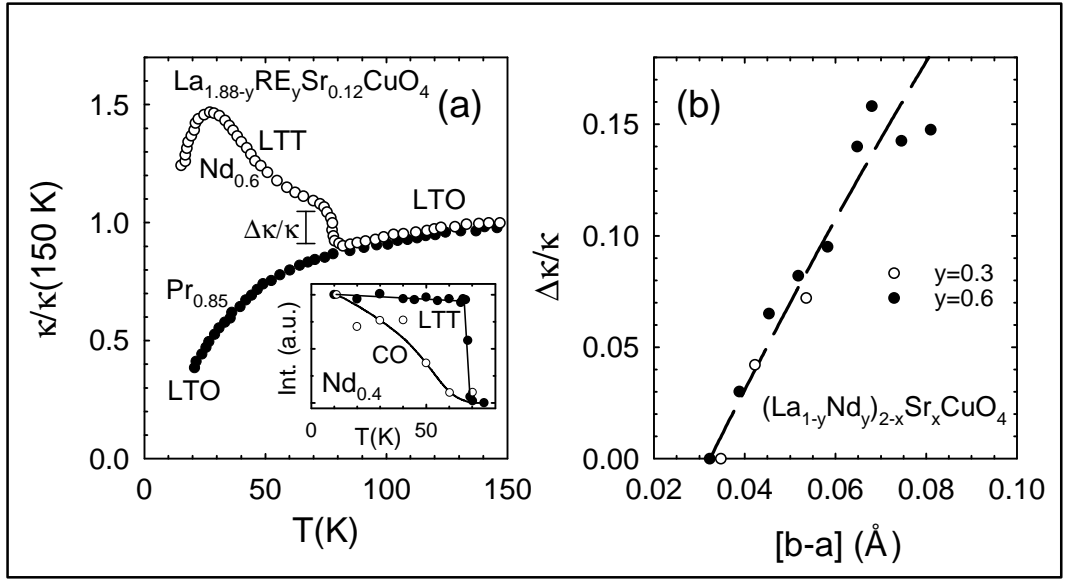

FIGURE 6 (a) Thermal conductivity *vs* temperature for RE-doped $La_{1.88}Sr_{0.12}CuO_4$ [16]. $\Delta\kappa/\kappa$ is the normalized jump in $\kappa$ at the LTO-LTT transition. The inset shows the integrated intensity of LTT and charge-order superlattice reflections of a $Nd_{0.4}$ single crystal [1]. (b) $\Delta\kappa/\kappa$ for Nd-doped LSCO [16] plotted against the orthorhombic splitting in the LTO phase.

tending to zero at the superconducting boundary. Using the structural data for Nd-doped compounds [15] we convert [17] Sr content to orthorhombic distortion, (b-a) in the LTO phase, and find that $\Delta\kappa/\kappa$ correlates well with (b-a) [Fig. 6 (b)]. Within the stripe scattering hypothesis, $\Delta\kappa/\kappa$ scales with the volume fraction of static stripe phase, and thus is larger for higher Nd content. Baberski *et al.* argued against any direct connection with the structural transition given that no anomaly occurred at the LTO-LTT transition when the LTT phase was superconducting (i.e. in the absence of static charge stripes). An apparent difficulty with this interpretation is that the charge-order develops rather gradually below $T_{LTT}$, in contrast to the $\kappa$ anomalies which are step-like jumps that follow closely the temperature behavior of the LTT structural order parameter [inset, Fig. 6 (a)]. Though the magnitude of the tilts is not altered at $T_{LTT}$, for a given tilt, the LTO-phase octahedra are more distorted than those of the LTT phase. Using diffraction data for $La_{1.475}Nd_{0.4}Sr_{0.125}CuO_4$ [18] we compute a decrease in the distortion parameter, $\Delta D \sim -0.8\%$ at the LTO-LTT transition. In the absence of a "calibration" of thermal resistivity *vs* distortion like that of Fig. 1 (b) for the manganites, we can only comment that such a change of distortion in the manganites would be more than sufficient to produce $\Delta\kappa/\kappa$ of the magnitude observed for $(La_{1-y}Nd_y)_xSr_xCuO_4$ [16]. Thus the plot in Fig. 6 (b) may be telling us that $\Delta\kappa/\kappa$ correlates with the change in distortion of the octahedra. Within this scenario, $\Delta\kappa/\kappa$ presumably tends to zero at a finite value of (b-a) because the corresponding tilt angle is sufficiently small that the difference in distortion for LTO and LTT tilts causes a negligible difference in phonon scattering. It may also be relevant to the doping behavior of $\Delta\kappa/\kappa$ that Sr dopants introduce localized holes in LSCO [19] that are associated with the presence of local LTT domains at T>$T_{LTT}$ [20]. As for the manganite and nickelate, there may be some hardening of the lattice that would also contribute to $\Delta\kappa/\kappa$, but we are not aware of sound velocity measurements for RE-doped LSCO.

## $YBa_2Cu_3O_{6+x}$ and Hg-cuprates

In the absence of a pinning mechanism like the tilt distortion of the LTT phase, there is no long-range, static charge ordering in the cuprates. However, inelastic neutron scattering studies of LSCO [21] and $YBa_2Cu_3O_{6+x}$ (Y-123) [22] suggest the presence of fluctuating or disordered charge-stripes. Our studies of the doping dependence of thermal conductivity in oxygen doped cuprates [23] reveal enhanced thermal resistivity near hole concentrations p=1/8, the value for which commensurate charge stripe order is expected. We attributed this to the presence of static stripe order in small domains.

The p=1/8 features (Fig. 7) are evident in the doping behavior of the normal-

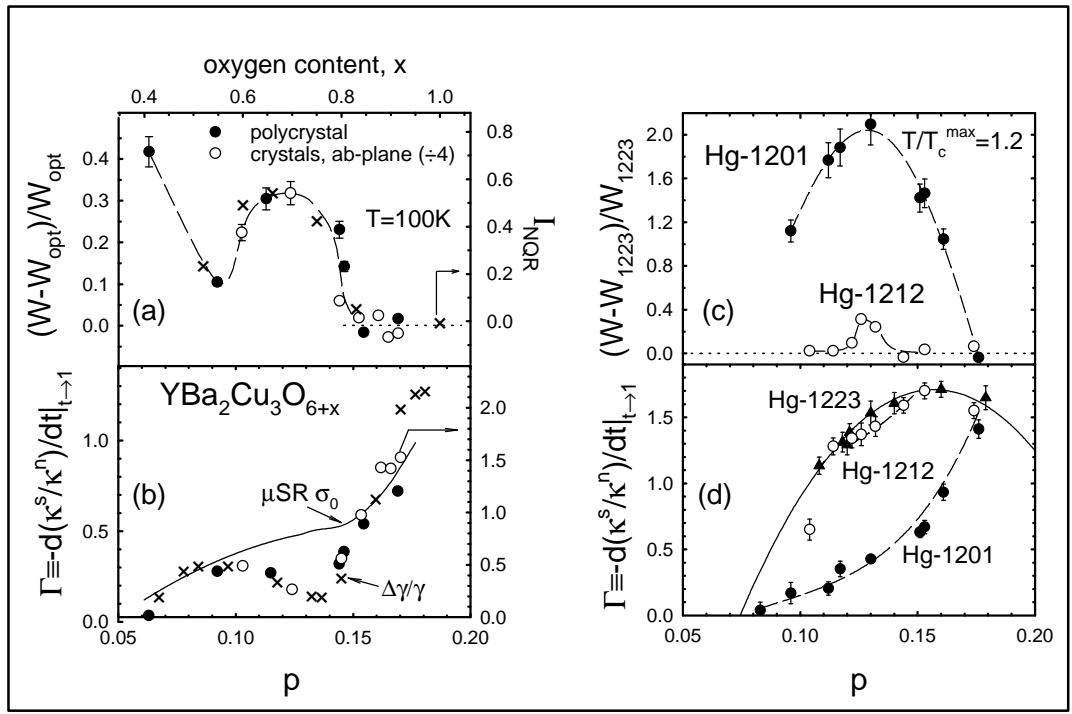

FIGURE 7 (a) Thermal resistivity [23] at T=100K relative to that at $p_{opt}$=0.16 for Y-123 polycrystals and the ab-plane of single crystals [$W_{opt}$=0.21 mK/W (0.08mK/W) for the polycrystal (crystals)]. Also plotted (×'s) is the relative intensity of anomalous $^{63}$Cu NQR signals [24]. The dashed curve is a guide to the eye. (b) The normalized slope change in $\kappa$(T) at $T_c$ vs doping for each of the Y-123 specimens in (a). Solid (open) symbols are referred to the left (right) ordinate. Also shown (×'s) are the normalized electronic specific heat jump [25], $\Delta\gamma/\gamma$, and (solid curve) the μSR depolarization rate [26] (in μs$^{-1}$), divided by 1.4 and 2.7, respectively, and referred to the right ordinate. (c) Thermal resistivity for Hg-1201 and Hg-1212 polycrystals relative to that of Hg-1223. Dashed curves are guides. (d) The normalized slope change in $\kappa$(T) at $T_c$ vs doping for Hg cuprates. The solid line is 1.71-250(p-0.157)$^2$. Dashed curves are guides.

state thermal resistivity (W) and the normalized change in temperature derivative of $\kappa$ that occurs at the superconducting transition ($T_c$), $\Gamma \equiv -d(\kappa^s/\kappa^n)/dt|_{t=1}$, where $t=T/T_c$ and $\kappa^s$ ($\kappa^n$) is the thermal conductivity in the superconducting (normal) state. For Y-123, W(p) and $\Gamma$(p) follow closely the doping behavior of anomalous $^{63}$Cu NQR spectral weight [24] and the electronic specific heat jump [25], $\Delta\gamma/\gamma$, respectively [crosses in Fig.'s 7 (a) and (b)]; this motivates our proposal that W probes lattice distortions associated with localized holes, and $\Gamma$ the change in low-energy spectral weight induced by superconductivity.

The muon spin rotation (μSR) depolarization rate [$\sigma_0$ in Fig. 7 (b)], proportional to the superfluid density, exhibits a smooth behavior through 1/8 doping [26]. The μSR signal originates in regions of the specimen where there is a flux lattice and this difference from the behavior of $\Gamma$ and $\Delta\gamma/\gamma$ (both bulk probes of the superconducting volume) suggests that the material is inhomogeneous, with non-superconducting inclusions. The suppression of $\Gamma$ and $\Delta\gamma/\gamma$ below the scaled $\sigma_0$ curve in Fig. 7 (b) are presumably measures of the non-superconducting volume fraction. Taken together the W and $\Gamma$ data imply that the non-superconducting regions are comprised of localized holes and associated lattice distortions, similar to what would be expected for stripe domains.

For the Hg materials the p=1/8 enhancement of W and suppression of $\Gamma$ is most prominent in single-layer Hg-1201, less so in double-layer Hg-1212, and absent or negligible in three-layer Hg-1223. This trend follows that of the oxygen vacancy concentration: a single HgO$_\delta$ layer per unit cell contributes charge to m planes in Hg-12(m-1)m so that the oxygen vacancy concentration, 1-$\delta$, increases with decreasing m [27]. The absence of suppression in $\Gamma$ near 1/8 doping for Hg-1223 [Fig. 7 (d)] suggests that this material has sufficiently few localized-hole domains that their effects in W and $\Gamma$ are unobservable. Thus we employ the Hg-1223 W(p) data as a reference and plot the differences for the other two compounds in Fig. 7 (c). Comparing Fig.'s 7 (c) and (d) we see that for both Hg-1201 and Hg-1212 W is enhanced and $\Gamma$ suppressed relative to values for Hg-1223 in common ranges of p, with maximal differences near p=1/8. As we have discussed elsewhere [23], these results suggest that clusters of oxygen vacancies play a role in localizing holes, possibly similar to that of the tilt distortions in LSCO.

## ACKNOWLEDGEMENTS

This work was supported by NSF Grant No. DMR-9631236.